\documentclass[10pt, a4, conference]{IEEEtran}
\IEEEoverridecommandlockouts
\usepackage[dvipsnames, svgnames, x11names]{xcolor}
\usepackage{amsmath,graphicx,amssymb,delarray,subfigure,verbatim,booktabs, amsfonts}
\usepackage{algorithmic, textcomp, xcolor, graphicx, cite}
\usepackage{bbding}
\usepackage{xfrac}
\usepackage{diagbox, makecell}
\usepackage{multirow}
\usepackage{colortbl}
\usepackage[english]{babel}
\usepackage{lipsum}

\def\BibTeX{{\rm B\kern-.05em{\sc i\kern-.025em b}\kern-.08em
    T\kern-.1667em\lower.7ex\hbox{E}\kern-.125emX}}
\begin{document}

\title{Learning to Inference with Early Exit in the Progressive Speech Enhancement}

\author{
\IEEEauthorblockN{Andong Li$^{\star \dagger}$, Chengshi Zheng$^{\star \dagger}$, Lu Zhang$^{\dagger\dagger}$ Xiaodong Li$^{\star \dagger}$}
\IEEEauthorblockA{$^{\star}$ Key Laboratory of Noise and Vibration Research, Institute of Acoustics, Chinese Academy\\
of Sciences, Beijing, China\\
$^{\dagger}$ University of Chinese Academy of Sciences, Beijing, China\\
$^{\dagger\dagger}$ Department of Electronics and Information Engineering, Harbin Institute of Technology, Shenzhen, China}
\{liandong, cszheng, lxd\}@mail.ioa.ac.cn, 18B952047@stu.hit.edu.cn

\thanks{This work was supported by National Key R\&D Program of China. This work was also supported by IACAS Young Elite Researcher Project under no.QNYC201813.}}

\maketitle
\begin{abstract}
In real scenarios, it is often necessary and significant to control the inference speed of speech enhancement systems under different conditions. To this end, we propose a stage-wise adaptive inference approach with early exit mechanism for progressive speech enhancement. Specifically, in each stage, once the spectral distance between adjacent stages lowers the empirically preset threshold, the inference will terminate and output the estimation, which can effectively accelerate the inference speed. To further improve the performance of existing speech enhancement systems, PL-CRN++ is proposed, which is an improved version over our preliminary work PL-CRN and combines stage recurrent mechanism and complex spectral mapping. Extensive experiments are conducted on the TIMIT corpus, the results demonstrate the superiority of our system over state-of-the-art baselines in terms of PESQ, ESTOI and DNSMOS. Moreover, by adjusting the threshold, we can easily control the inference efficiency while sustaining the system performance.   
\end{abstract}

\section{Introduction}
\label{sec:intro}
Speech enhancement (SE) aims to extract the target speech signals from the  corrupted noisy mixtures, facilitating the application of speech techniques in real scenarios, like telecommunication systems and hearing assistant devices~{\cite{loizou2013speech}. Thanks to the development of deep neural networks (DNNs), the performance of SE systems are notably boosted in recent years with the help of complicated network topology, including feedforward layers~{\cite{xu2014regression}}, convolutional neural networks (CNNs)~{\cite{park2017fully, multi2020}}, long-short term memory units (LSTMs)~{\cite{chen2017long}}, and self-attention mechanism~{\cite{koizumi2020speech}}.
	
Motivated by curriculum learning concept~{\cite{bengio2009curriculum}}, progressive learning (PL) based SE algorithms begin to thrive recently~{\cite{gao2018densely, li2020speech, li2020recursive}}, which have been demonstrated to exhibit better performance over the conventional ``black-box'' DNN framework. Different from the previous works where the whole mapping process is viewed as agnostic, PL explicitly decomposes the original difficult task into several easier sub-problems and the speech can be recovered in a stage-wise manner. Despite the advantages of multi-stage based SE algorithms, they usually suffer from the ``\textbf{heavy run-time delay}'' problem, which can be elaborated from two perspectives. On the one hand, for real-time devices, the processing delay needs to be considered seriously. However, the number of network depth will linearly grow with the increase of training stages, which usually brings linear-proportional processing delay~{\cite{li2020speech}}. On the other hand, as the inference of current stage usually concerns the estimation results from previous stages, the system has to wait until the previous stages are finished, which also heavily limits the parallelism of SE systems.

To mitigate the problem above, an \textbf{E}arly \textbf{E}xit \textbf{M}echanism named EEM has been applied into pretrained language model (PLM)~{\cite{zhou2020bert}} and multi-channel source separation (MSS)~{\cite{chen2020don}}. Inspired by this idea, we propose a new EEM method for adaptive inference with more fast and robust speech enhancement performance. Specifically, a pre-defined threshold is set beforehand. During the inference period, the calculation will not be stopped until the estimation gap between the adjacent stages is lower than the threshold. In this way, the model can early exit without passing through all layers. As the algorithm only works in the inference stage, it is simple to operate and can be generalized to various SE systems. The advantages of the proposed EEM can be illustrated as two-fold. Firstly, we can adaptively switch the output result depending on the device requirement. For example, for the devices with strict run-time delay requirement, faster inference time can be achieved by setting suitable threshold values. Secondly, we can also adaptively control the inference time depending on the SNR requirement. For example, in low SNRs, we can pass more stages to generate more clear speech while fewer stages are needed in high SNRs.
	
Recently, the benefit of phase recovery in improving speech perception especially in low SNRs has been well investigated~{\cite{tan2019learning, pandey2019new, li2020time, yin2020phasen}}. In our preliminary study~{\cite{li2020speech}}, the PL-CRN only works for magnitude enhancement. In this paper, we further propose an improved PL-CRN called PL-CRN++. Several strategies are adopted. Firstly, instead of estimating magnitude, both real and imaginary (RI) components of the spectrum are estimated simultaneously~{\cite{tan2019learning}}. Besides, all the (de)convolution layers in the encoder and decoder modules are replaced by the gated linear unit (GLU) formats~{\cite{dauphin2017language}}. In addition, rather than simply concatenate the outputs from previous stages along the channel dimension as the input of current stage, we adopt a stage recurrent mechanism to effectively grasp the sequence dependency across different stages~{\cite{li2020time}}. 
	
The rest of the paper is organized as follows. In Section~{\ref{sec:proposed-approach}}, both early exit mechanism and the utilized network are introduced. In Section~{\ref{sec:experimental-setup}}, the experimental settings are given. Experimental results and analysis are presented in Section~{\ref{sec:results-and-analysis}}. Some conclusions are drawn in Section~{\ref{sec:conclusion}}. 
	
	\begin{figure*}[t]
		\centering
		\centerline{\includegraphics[width=1.8\columnwidth]{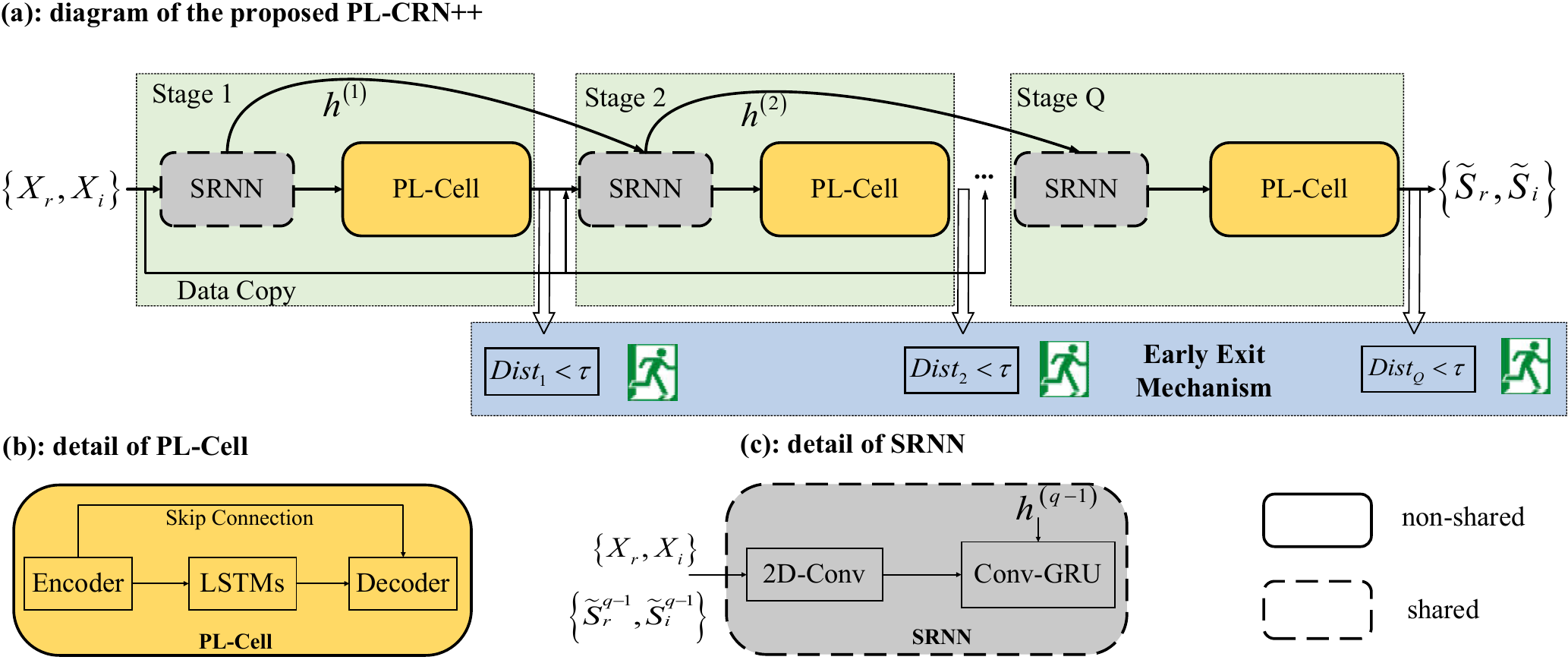}}
		\caption{Proposed PL-CRN++ incorporating early exit mechanism. (a): overall flowchart of the proposed approach. (b): the pipeline detail of the PL-Cell. (c): the pipeline detail of SRNN. }
		\label{fig:architecture}
		\vspace{-0.4cm}
	\end{figure*}

\section{Proposed Approach}
\label{sec:proposed-approach}
\subsection{Progressive Learning for Speech Enhancement}
\label{sec:progressive-learning}
With short-time Fourier transform (STFT), let $X\left(k,l\right)$, $S\left(k,l\right)$, and $N\left(k,l\right)$ denote the complex values of noisy, clean, and noise signals in the T-F domain with the frequency index of $k$ and time index of $l$. The problem is formulated as:
\begin{equation}
\label{eqn1}
X\left(k,l\right) = S\left(k,l\right) + N\left(k,l\right), k\in[1,K], l\in[1,L],
\end{equation}
where $K$ and $L$ denote the number of frequency bins and time frames, respectively. For better readability, we will omit the T-F index if no confusion arises. In the progressive learning based speech enhancement approaches, the whole denoising process is split into several stages, and the intermediate targets are defined as the SNR-improved spectra over the noisy version. Let $Q$ denote the total stage number, then the intermediate targets are defined as $\left\{S^{1}, S^{2}, \cdots, S^{Q} \right\}$. In the $q_{th}$ stage, the mapping process can be presented as:
\begin{equation}
\label{eqn2}
\tilde{S}^{q} = \mathcal{G}_{q}\left( X, \tilde{S}^{1},\cdots, \tilde{S}^{q-1}; \Theta_{q} \right),
\end{equation}
where $\mathcal{G}_{q}\left(\cdot\right)$ and $\Theta_{q}$ denote the mapping function and the parameter set in the $q_{th}$ stage, respectively. $\tilde{S}^{q}$ refers to the estimated complex spectrum in the $q_{th}$ stage. To enable the network training, we concatenate the real and imaginary components of the spectrum along the channel axis, \emph{i.e.}, $\left\{ X, \tilde{S}^{1}, \cdots \tilde{S}^{Q} \right\} \in \mathbb{R}^{2\times K \times L}$. Note that in the current stage, the previous outputs together with the noisy complex spectrum are involved as the inputs to mitigate the information loss during the training process~{\cite{gao2018densely}}.

\subsection{Early Exit Mechanism}
\label{sec:early-exit}
Early exit mechanism (EEM) is originally proposed in~{\cite{kaya2019shallow}} to mitigate the ``overthinking'' problem during the decision-making process. That is, for many input samples, shallow representation is already adequate for network classification. In this paper, we rethink this problem from another perspective. In real scenarios, the requirement for inference speed varies among different devices. Besides, noise intensity in the real environment usually changes largely. When the SNR is quite low, it is necessary to pass more stages to generate the speech with better quality. However, for high SNR cases, fewer stages are needed to yield the enhanced speech with adequate quality.

Informally, we define $\tau$ as the threshold, which is manually set beforehand. For $q_{th}$ stage, the adjacent spectral distance $Dist_{q}$ is defined as:
\begin{equation}
\label{eq3}
Dist_{q} \triangleq \frac{1}{ZLK}\sum_{l=1}^{L}\sum_{k=1}^{K}       
  \left\|\tilde{S}^{q}\left(k,l\right), \tilde{S}^{q-1}\left(k,l\right)\right\|_{2}^{2},
\end{equation}
The normalization term $Z$ is set to mitigate the influence of magnitude scale, defined as follows.
\begin{equation}
\label{eq4}
Z \triangleq \frac{1}{LK}\sum_{l=1}^{L}\sum_{k=1}^{K}\left\| X(k, l) \right\|^{2}_{2},
\end{equation}
Note that for $q=1$ case, the distance is calculated between the estimation in the first stage $\tilde{S}^{1}$ and the noisy version, \emph{i.e.}, $\tilde{S}^{0} \equiv X$.
 
As shown in Figure~{\ref{fig:architecture}}(a), during the run time, in each stage, the estimation gap needs to be calculated. If $Dist_{q} < \tau$, we will terminate the inference and the estimated spectrum in the current stage will be chosen as the final enhanced result. We argue that the dominant noise components tend to be suppressed in the first several stages, leading to the gradual decrease of the spectral distance in the latter. Therefore, a larger threshold value will result in faster inference termination accordingly, and vice versa. The experiments are conducted in Sec.~{\ref{sec:results-and-analysis}} to support the point. As a result, we can dynamically control the inference efficiency by setting different thresholds empirically. 
\subsection{PL-CRN++}
\label{sec:proposed-framework}
The proposed framework is presented in Figure~{\ref{fig:architecture}}(a), which is composed of multiple stages. Within each stage, the stage-recurrent neural network (SRNN) is adopted to learn the sequence information across stages~{\cite{li2020time}}. PL-Cell takes a typical ``Encoder-LSTM-Decoder'' topology~{\cite{braun2021towards}} and is tasked with mapping to the specific intermediate target. The overall paradigm is similar to PL-CRN~{\cite{li2020speech}}, except some improvements are provided. Firstly, instead of estimating the spectral magnitude, we take the complex spectral mapping strategy, \emph{i.e.}, RI components serve as the input and target. In this way, magnitude and phase can be recovered simultaneously, which is beneficial to speech quality improvement. Secondly, similar to~{\cite{tan2019learning}}, within each PL-Cell, we replace all the (de)convolution layers in the encoder and decoder with the gated linear unit (GLU) formats~{\cite{dauphin2017language}}, where another convolutional branch with a sigmoid activation function is introduced to recalibrate the feature distribution in the major branch. Thirdly, motivated by~{\cite{li2020time}}, we adopt the SRNN to establish the sequence dependency across different stages, which is shown in Figure~{\ref{fig:architecture}}(c). 

In each stage, the forward calculation process can be formulated as:
\begin{gather}
\label{eq5}
\hat{h}^{q} = f_{2d\_conv}\left(Cat[X_{r}, X_{i}, \tilde{S}^{q-1}_{r}, \tilde{S}^{q-1}_{i}]\right),\\
h^{(q)} = f_{convgru}\left(\hat{h}^{q}, h^{(q-1)}\right),\\
\tilde{S}^{q} = f_{decoder}\left( f_{lstm}\left( f_{encoder}\left( h^{(q)}     \right)\right) \right),
\end{gather}
where $f_{2d\_conv}$, and $f_{convgru}$ denote the functions of two-dimensional (2-D) convolution and Conv-GRU~{\cite{ballas2015delving}} in the SRNN. $f_{encoder}$, $f_{lstm}$, and $f_{decoder}$ are the functions of the encoder, LSTM and decoder in the PL-Cell, respectively. $Cat[\cdot]$ denotes the catenation operation along the channel axis.

Detailed network parameters are configured below. For all the 2-D convolutions in PL-CRN++, the number of channels is 64 except the output, where the channel is set to 2 to generate the real and imaginary parts. The kernel size and the stride are (2, 3) and (1, 2) along the time and frequency axis, respectively. To facilitate training convergence, after each 2-D convolution, instance normalization (IN)~{\cite{ulyanov2016instance}} and Parameter ReLU (PReLU)~{\cite{he2015delving}} are followed. For the encoder module, five consecutive convolutional blocks are utilized to compress the feature size while the decoder is the mirror version of the encoder to reconstruct the spectrum. Two LSTM layers with 256 units are utilized as the sequence modeling module. Note that, the parameter weights in each PL-Cell are not shared while SRNN is reused in each stage. 

\subsection{Loss Function}
Similar to~{\cite{chen2020don, kaya2019shallow}}, we take the following weighted loss for network training:
\begin{equation}
\label{eqn6}
\mathcal{L} = \frac{1}{\sum_{q=1}^{Q}} \sum_{q=1}^{Q}q\left\| \tilde{S}^{q} - S^{q}  \right\|_{2}^{2},
\end{equation} 
The behind rationale lies in that with the increase of stage index, a larger loss will be given, corresponding to the relative inference cost of each intermediate stage.

\section{Experimental Setup}
\label{sec:experimental-setup}
\subsection{Dataset}
In this study, we conduct the experiments on the TIMIT corpus~{\cite{garofolo1993darpa}}. 4620, 400, and 150 utterances are utilized for training, validation, and testing, respectively. No utterance overlap exists among the three parts. For noise-robust training, around 20,000 noises are randomly selected from the DNS-Challenge{\footnote{https://github.com/microsoft/DNS-Challenge}} to obtain a 55 hours noise set for training. To create multiple SNR-improved intermediate targets, we fix the total stage number $Q$ as 5. During each mixed process, a random cut is generated to obtain a noise vector, which is subsequently mixed with a randomly chosen clean utterance. The SNR range for training is [-5\rm{dB}, 30\rm{dB}] with 2dB interval. After an SNR value is randomly selected, we improve the SNR by 10dB after each stage and the target in the final stage is the clean version. Totally, 100,000 pairs are created for training (around 85 hours).

For model evaluation, four challenging unseen noises are selected, where babble, factory1, white from NOISEX92~{\cite{varga1993assessment}}, and cafeteria noise from the CHIME3 dataset~{\cite{barker2015third}}. Four SNR conditions are explored, namely -5\rm{dB}, 0\rm{dB}, 5\rm{dB}, and 10\rm{dB}. 150 pairs are generated for each case.

\renewcommand\arraystretch{1.25}
\begin{table*}[t]
	\caption{Evaluation results among different models in terms of PESQ, ESTOI and DNSMOS for different SNRs. Comparison w.r.t. speed-up ratio is also presented among various thresholds.}
	\Huge
	\centering
	\resizebox{\textwidth}{!}{
		\begin{tabular}{c|c|c|c|c|ccccc|ccccc|ccccc|ccccc}
			\toprule
			&\textbf{Metrics}
			&\multirow{2}*{\rotatebox{90}{\textbf{Gate}}} &\multirow{2}*{\rotatebox{90}{\textbf{SRNN}}}  &\multirow{2}*{{\textbf{\#Param}}} &\multicolumn{5}{c|}{\textbf{Speed-up ratio}} &\multicolumn{5}{c|}{\textbf{PESQ}} &\multicolumn{5}{c|}{\textbf{ESTOI(\%)}} &\multicolumn{5}{c}{\textbf{DNSMOS}} \\
			\cline{2-2}\cline{6-25}
			&\textbf{SNR(dB)} & & & &-5 &0 &5 &10 &Avg. &-5 &0 &5 &10 &\multicolumn{1}{c|}{Avg.} &-5 &0 &5 &10 &\multicolumn{1}{c|}{Avg.} &-5 &0 &5 &10 &\multicolumn{1}{c}{Avg.}\\
			\cline{1-25}
			&\multicolumn{1}{c|}{\textbf{Noisy}} &-  &- &- &- &- &- &- &- &1.34 &1.69 &2.07 &2.43 &\multicolumn{1}{c|}{1.88} &29.49 &42.51 &57.40 &71.50 &\multicolumn{1}{c|}{50.23} &2.22 &2.34 &2.57 &2.78 &\multicolumn{1}{c}{2.48}\\
			\cline{1-25}
			\multirow{7}*{\rotatebox{90}{\textbf{Model Comparison}}}
			&\textbf{PL-LSTM} &- &- &38.13M &- &- &- &- &- &1.67 &2.05 &2.43 &2.77 &2.23 &36.44 &51.14 &65.28 &75.28 &57.03 &2.51 &2.63 &2.82 &3.02 &2.75\\
			&\multicolumn{1}{c|}{\textbf{PL-CRN}} &- &- &5.55M &- &- &- &- &- &1.82 &2.21 &2.61 &2.98 &2.40 &41.54 &57.04 &71.81 &82.27 &63.17 &2.65 &2.79 &2.99 &3.22 &2.91\\
			&\multicolumn{1}{c|}{\textbf{GCRN}} &- &- &18.16M &- &- &- &- &- &1.88 &2.37 &2.82 &3.16 &2.56 &51.30 &68.69 &81.06 &88.18 &72.31 &2.88 &3.11 &3.35 &3.56 &3.23 \\
			\cline{2-25}
			&\multirow{4}*{\textbf{PL-CRN++}} &\XSolidBrush &\XSolidBrush &7.18M &- &- &- &- &- &1.81 &2.30 &2.77 &3.19 &2.52 &51.38 &68.68 &81.34 &88.76 &72.54 &3.06 &3.31 &3.54 &3.72 &3.40\\
			& &\Checkmark &\XSolidBrush &9.09M &- &- &- &- &-  &1.95 &2.39 &2.83 &3.23 &2.60 &53.38 &69.98 &82.17 &89.10 &73.66 &3.10 &3.34 &3.56 &3.73 &3.43\\
			& &\XSolidBrush &\Checkmark  &7.52M &- &- &- &- &- &1.83 &2.31 &2.78 &3.19 &2.53 &53.80 &70.46 &82.71 &89.54 &74.13 &3.12 &3.37 &3.56 &3.75 &3.45\\
			& &\Checkmark &\Checkmark &9.61M &- &- &- &- &-  &1.94 &2.44 &2.87 &3.26 &2.63 &56.35 &72.55 &83.86 &90.23 &75.75 &3.16 &3.43 &3.65 &3.81 &3.51\\
			\midrule
			\multirow{7}*{\rotatebox{90}{\textbf{Early Exit}}} &\multicolumn{1}{c|}{\textbf{PL-CRN++ ($\tau=+\infty$)}} &\Checkmark &\Checkmark &9.61M &5.00$\times$ &5.00$\times$ &5.00$\times$ &5.00$\times$ &5.00$\times$  &1.66 &2.05 &2.43 &2.80 &2.23 &35.27 &50.59 &67.01 &80.09 &58.24 &2.48 &2.68 &2.88 &3.13 &2.79\\
			&\multicolumn{1}{c|}{\textbf{PL-CRN++ ($\tau=0.6$)}} &\Checkmark &\Checkmark &9.61M &2.87$\times$ &4.46$\times$ &5.00$\times$ &5.00$\times$ &4.11$\times$  &1.85 &2.08 &2.43 &2.80 &2.29 &42.01 &52.11 &67.01 &80.09 &60.30 &2.65 &2.71 &2.88 &3.13 &2.84\\
			&\multicolumn{1}{c|}{\textbf{PL-CRN++ ($\tau=0.2$)}} &\Checkmark &\Checkmark &9.61M &2.55$\times$ &2.92$\times$ &4.81$\times$ &5.00$\times$ &3.50$\times$  &1.91 &2.29 &2.64 &2.80 &2.41 &44.35 &60.85 &75.39 &80.13 &65.18 &2.72 &2.91 &3.14 &3.14 &2.98\\
			&\multicolumn{1}{c|}{\textbf{PL-CRN++ ($\tau=0.08$)}} &\Checkmark &\Checkmark &9.61M &2.46$\times$ &2.59$\times$ &2.74$\times$ &4.97$\times$ &2.94$\times$ &2.02 &2.43 &2.75 &3.02 &2.55 &50.15 &66.90 &78.61 &85.99 &70.41 &2.95 &3.18 &3.28 &3.46 &3.22\\
			&\multicolumn{1}{c|}{\textbf{PL-CRN++ ($\tau=0.04$)}} &\Checkmark &\Checkmark &9.61M &1.71$\times$ &1.77$\times$ &2.09$\times$ &2.58$\times$ &1.98$\times$ &2.02 &2.45 &2.83 &3.12 &2.60 &52.29 &68.69 &81.00 &87.83 &72.45 &3.00 &3.23 &3.43 &3.59 &3.31\\
			&\multicolumn{1}{c|}{\textbf{PL-CRN++ ($\tau=0.02$)}} &\Checkmark &\Checkmark &9.61M &1.46$\times$ &1.53$\times$ &1.69$\times$ &1.95$\times$ &1.64$\times$ &1.99 &2.46 &2.86 &3.19 &2.63 &54.70 &71.17 &82.66 &88.94 &74.37 &3.08 &3.31 &3.51 &3.68 &3.40\\
			&\multicolumn{1}{c|}{\textbf{PL-CRN++ ($\mathbf\tau=0.01$)}} &\Checkmark &\Checkmark &9.61M &1.20$\times$ &1.20$\times$ &1.31$\times$ &1.58$\times$ &1.31$\times$ &1.95 &2.44 &2.86 &3.24 &2.62 &55.90 &72.20 &83.61 &89.79 &75.38 &3.14 &3.39 &3.61 &3.74 &3.47\\
			&\multicolumn{1}{c|}{\textbf{PL-CRN++ ($\mathbf\tau=0$)}} &\Checkmark &\Checkmark &9.61M &1.00$\times$ &1.00$\times$ &1.00$\times$ &1.00$\times$ &1.00$\times$ &1.94 &2.44 &2.87 &3.26 &2.63 &56.35 &72.55 &83.83 &90.23 &75.75 &3.16 &3.43 &3.65 &3.81 &3.51\\
			\bottomrule
	\end{tabular}}
	\label{tbl:evaluation-results}
	\vspace{-0.4cm}
\end{table*}

\subsection{Parameter Configurations}
\label{sec:parameter-configurations}
All the utterances are sampled at 16kHz. The window size is 20ms with 50\% overlap in adjacent frames. 320-point FFT is utilized to extract 161-D spectral features. As complex spectral mapping strategy is adopted, we concatenate the RI along channel axis, $i.e.$, $Cat[X_{r}, X_{i}] \in \mathbb{R}^{2\times K \times L}$. The model is optimized by Adam~{\cite{kingma2014adam}}. The learning rate is initialized at 1e-3, which will be halved if consecutive 3 loss increases arise. The total epoch number is set to 50 with the batch number being 8.

To analyze the impact of early exit mechanism, we set multiple $\tau$ candidates, including $\{+\infty, 0.6, 0.2, 0.08, 0.04, 0.02, 0.01, 0\}$. We also evaluate the performance with another three advanced baseline systems, namely PL-LSTM~{\cite{gao2018densely}}}, PL-CRN~{\cite{li2020speech}}, and GCRN~{\cite{tan2019learning}}. Both PL-LSTM and PL-CRN belong to the PL family and we also fix the stage number $Q$ as 5. In~{\cite{li2020speech}}, the LSTMs within PL-CRN are shared across different stages and we cancel the shared option in this study to boost the overall performance. Note that, all the models are causal-designed for fair comparison.

\section{Results and Analysis}
\label{sec:results-and-analysis}
Two metrics are utilized to evaluate the objective performance of different systems, namely perceptual evaluation of speech quality (PESQ)~{\cite{rix2001perceptual}}, and extended short-time objective intelligibility (ESTOI)~{\cite{jensen2016algorithm}}. Besides, to evaluate the subjective quality, DNSMOS is also adopted, which is a robust non-intrusive speech quality metric and well suitable for accurate subjective rating~{\cite{reddy2020dnsmos}}.

\subsection{Ablation Study}
\label{ablation-study}
We investigate the effect of SRNN and gating branch in the GLU, whose results are shown in the middle region of Table.~{\ref{tbl:evaluation-results}}. One can find that both two modules can effectively improve the metric performance. For example, compared with the naive PL-CRN++ (no gate and no SRNN), around 0.08 and 0.01 PESQ improvements are provided if SRNN and gating branch are applied, respectively. For DNSMOS, the similar tendency can be observed. Moreover, when both two modules are applied, the performance of PL-CRN++ can be further improved.  

\subsection{Metric Comparison Among Different Systems}
\label{sec:model-comparison}
We then compare our PL-CRN++ with another three baseline systems, as shown in the top region of Table~{\ref{tbl:evaluation-results}}. Several observations can be made. Firstly, our PL-CRN++ notably outperforms previous PL based systems, \emph{i.e.}, PL-LSTM and PL-CRN. For example, in terms of PESQ, our system yields around 0.40 and 0.23 improvements over PL-LSTM and PL-CRN. A similar trend is also observed for ESTOI. It indicates that by incorporating the stage recurrent mechanism and complex spectral mapping, we can make further breakthrough over current PL based algorithms. Secondly, compared with GCRN, a state-of-the-art SE system with complex spectral mapping, our system still yields consistent advantages. one can observe that our system achieves around 0.07 and 3.44\% improvements in terms of PESQ and ESTOI, respectively, which reveals the superiority of our approach. Thirdly, in terms of subjective quality, our approach sizably surpasses previous systems. For DNSMOS, our approach yields 0.76, 0.60, and 0.28 improvements over PL-LSTM, PL-CRN, and GCRN, respectively. It shows that our system can dramatically improve the subjective quality of enhanced speech, which is quite beneficial to speech perception under noisy conditions.

\subsection{The Effect of Early Exit Mechanism}
\label{impact-of-eem}
We adopt the speed-up ratio~{\cite{liu2020fastbert}} as the criterion to analyze the effect of early exit mechanism{\footnote{Within each stage, the same forward stream is adopted, leading to the same FLOPs. So speed-up ratio can be approximated as the ratio between the total stage index and the real stage index induced by early exit mechanism.}}, and a larger value indicates faster inference time. From the results in the bottom region of Table~{\ref{tbl:evaluation-results}}, several interesting observations can be made. Firstly, the decrease of $\tau$ will bring a smaller speed-up ratio, \emph{i.e.}, more inference cost will arise. This is because the inference will terminate as long as the adjacent spectral distance $Dist_{q}$ lowers the threshold. Therefore, larger as $\tau$ is, easier for the system to exit, and vice versa. Note that $\tau=+\infty$ and $\tau=0$ are two special cases, where the inference will terminate at the end of the first stage anyway for the former, and all the stages have to be passed for the latter. Secondly, for a fixed $\tau$, the increase of the input SNR will bring a larger speed-up ratio. We argue that for low SNR cases, noise components usually dominate the spectrum, so the adjacent spectral distance is relatively larger, and more stages are needed to meet the threshold. However, in the high SNRs, fewer noise components exist and can be easily removed at the early stage.

\begin{figure}[t]
	\centering
	\centerline{\includegraphics[width=0.85\columnwidth]{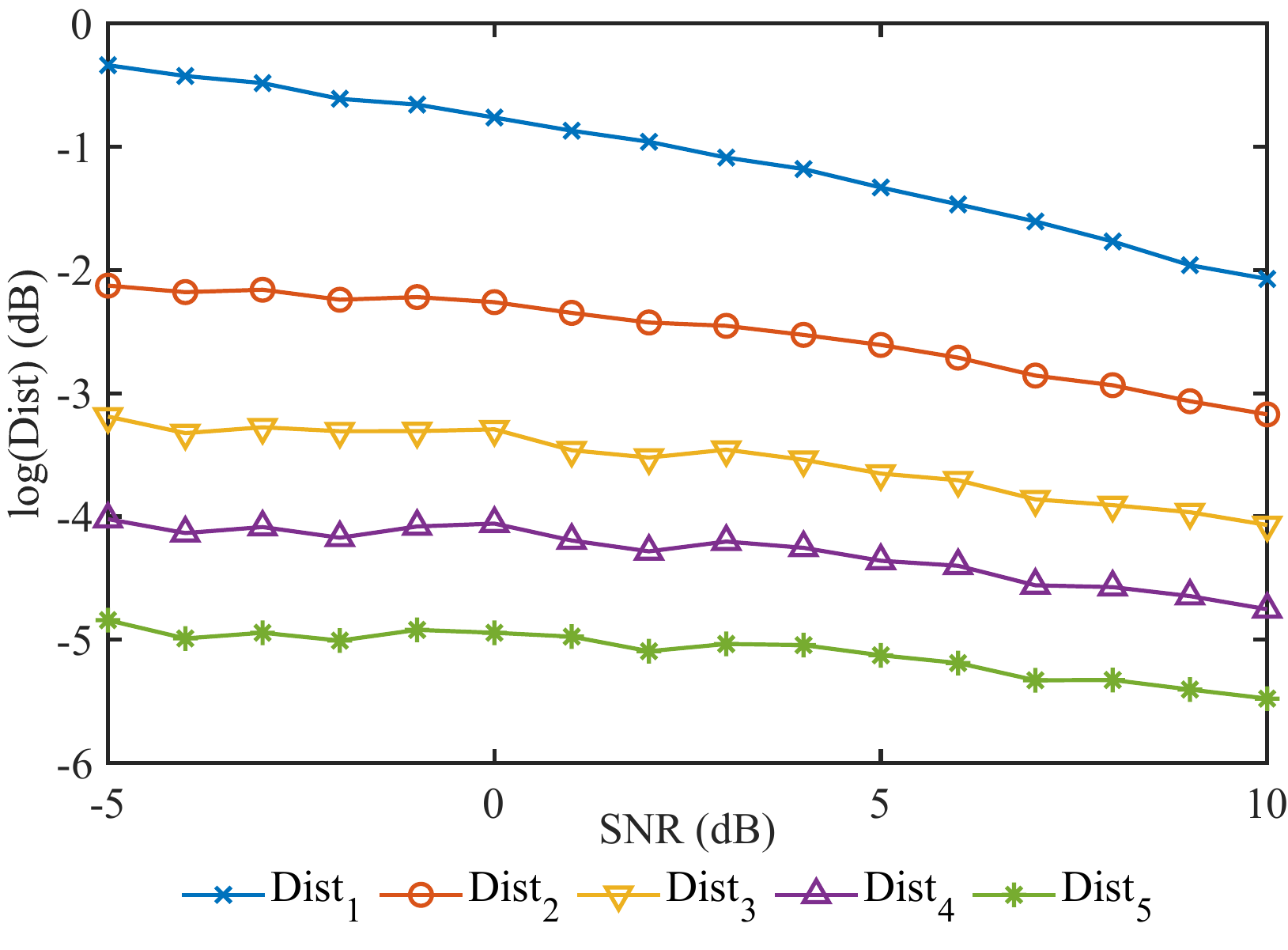}}
	\caption{Adjacent spectral distance $Dist_{i}$ under different SNRs. logarithm operation is adopted for better visualization.}
	\label{fig:dist-snr}
	\vspace{-0.4cm}
\end{figure}

Figure~{\ref{fig:dist-snr}} presents the $Dist_{q}$ at different stages under different input SNR conditions. Logarithm scale is adopted for better visualization. One can find that the distance will gradually decrease with the increase of stage index, indicating that in the PL based enhancement algorithms, more noises tend to be removed in the early stage and the late stages are mainly tasked with minor spectral refinement, which also follows the ``from coarse to fine" logic. Besides, higher SNRs will lead to smaller $Dist_{q}$, which also validates our point that a higher SNR brings faster inference time with early exit mechanism. 

\section{Conclusion}
\label{sec:conclusion}
We propose a stage-wise adaptive inference approach with early exit mechanism called EEM for fast and robust progressive speech enhancement. Specifically, a threshold is empirically set beforehand. During the run time, the adjacent spectral distance is calculated at each intermediate stage. Once it lowers the threshold, the procedure will terminate and output the enhanced result with early exit. To improve the performance of existing PL based systems, we propose PL-CRN++, which incorporates stage recurrent mechanism and complex spectral mapping strategy. The experiments on the TIMIT corpus show that PL-CRN++ consistently surpasses state-of-the-art baseline systems in multiple evaluation metrics. Moreover, by suitably selecting the threshold, EEM can adaptively control the inference speed of the model to perform more efficiently in real scenarios.

\end{document}